\documentclass[12 pt, amsfonts,amsmath, amssymb,color]{article}
\usepackage{authblk}
\def\baselinestretch{1.0}
\usepackage{amsmath,amssymb,amsfonts,latexsym,float,graphics,epsfig}
\usepackage{subfig}
\usepackage{verbatim}
\usepackage{relsize}
\usepackage{hyperref}
\usepackage{graphicx}
\usepackage{bm}
\usepackage{epstopdf}
\usepackage{slashed}
\usepackage{color}

\def\be{\begin{equation}}
\def\ee{\end{equation}}
\def\bea{\begin{eqnarray}}
\def\eea{\end{eqnarray}}

\begin{document}
%<<<<<<<<<<< enumeration of eqns section wise>>>>>>>>>>>>>>>>>>>

\renewcommand\theequation{\arabic{section}.\arabic{equation}}
\catcode`@=11 \@addtoreset{equation}{section}
%<<<<<<<<<<<<<<<<<<<<<<<<<<<<<<<<<>>>>>>>>>>>>>>>>>>>>>>>>>>>>>>>>>
\newtheorem{axiom}{Definition}[section]
\newtheorem{theorem}{Theorem}[section]
\newtheorem{axiom2}{Example}[section]
\newtheorem{lem}{Lemma}[section]
\newtheorem{prop}{Proposition}[section]
\newtheorem{cor}{Corollary}[section]

\newcommand{\ben}{\begin{equation*}}
\newcommand{\een}{\end{equation*}}

%%%%%%%%%%%

%\baselineskip 24pt

%%%%%%%%%%%
\let\endtitlepage\relax

\begin{titlepage}
\begin{center}
\renewcommand{\baselinestretch}{1.5}  %Line spacing
%\setstretch{1.5}

\vspace*{-0.5cm}

{\fontsize{19pt}{22pt}\bf{Logarithmic corrections to the entropy function of black holes in the open ensemble}}

\vspace{9mm}
\renewcommand{\baselinestretch}{1}  %Line spacing
%\setstretch{1}

\centerline{\large{Aritra Ghosh$^{\ddagger}$}\footnote{ag34@iitbbs.ac.in}, \large{Sudipta Mukherji$^{\dagger}$}\footnote{mukherji@iopb.res.in} and \large{Chandrasekhar Bhamidipati$^\ddagger$}\footnote{chandrasekhar@iitbbs.ac.in}}

\vspace{5mm}
\normalsize
\textit{$^\ddagger$School of Basic Sciences, Indian Institute of Technology Bhubaneswar, Jatni, Khurda, Odisha, 752050, India}\\
\textit{$^\dagger$Institute of Physics, Sachivalaya Marg, Bhubaneswar, 751005, India}\\
\textit{$^\dagger$Homi Bhabha National Institute, Training School Complex, Anushakti Nagar, Mumbai, 400085, India}\\
\vspace{5mm}

%--------------------------------------------------------------------------------------------------------------------------------------
\begin{abstract}
An `open' or $(\mu,P,T)$-ensemble describes equilibrium systems whose control parameters are chemical potential $\mu$, pressure $P$ and temperature $T$. Such an unconstrained ensemble is seldom used for applications to  standard thermodynamic systems due to the fact that the corresponding free energy identically vanishes as a result of the Euler relation. However, an open ensemble is perfectly regular for the case of black holes, as the entropy is a quasi-homogeneous function of extensive thermodynamic variables with scaling dictated by the Smarr formula. Following a brief discussion on thermodynamics in the open ensemble, we
compute the general form of logarithmic corrections to the entropy of a typical system, due to fluctuations in energy, thermodynamic volume and a generic charge $N$. This is then used to obtain the exact analytic form of the logarithmically corrected black hole entropy for charged and rotating black holes in anti-de Sitter spacetimes.
\end{abstract}
\end{center}
\vspace*{0cm}

% \renewcommand{\baselinestretch}{0.2}
% \setlength{\baselineskip}{0.1\baselineskip}
% \setstretch{0.1}

 %\tableofcontents
\end{titlepage}
\vspace*{0cm}
%%%%%%%5

\section{Introduction}
The association of the notion of a temperature and an entropy with a black hole horizon points towards the existence of a microscopic description of gravitational systems \cite{Bekenstein:1973ur}-\cite{Hawking:1974sw}. Following Boltzmann's insight: ``If you can heat it up, it has a microscopic structure", it is natural to expect that the thermodynamic properties of black holes can perhaps be derived from a microscopic theory of quantum gravity. Although the framework of classical thermodynamics is built on a macroscopic footing without necessarily any direct reference to an underlying microscopic theory, the formalism is robust enough to allow one to deduce useful insights about the microscopic framework. Perhaps the most insightful mathematical relationship connecting the macroscopic and microscopic formalisms is the Boltzmann's formula given by\footnote{We set \(k_B =1\) in subsequent discussions.},
\begin{equation}
  S = \ln \Omega_m
\end{equation} where \(\Omega_m\) is the number of microstates. It is speculated that there are microscopic gravitational degrees of freedom which account for the black hole entropy, which need to be understood. There has been a considerable amount of progress in this direction from the point of view of string theory, particularly for extremal black holes (see e.g., \cite{Vafa}-\cite{Mathur}). Remarkably, for several such systems the black hole entropy computed from counting string states agrees with the macroscopically defined Bekenstein-Hawking entropy. Interestingly, the black hole entropy receives perturbative logarithmic corrections both within general string and gravity models \cite{9407001}-\cite{Pourhassan:2015cga}, as well as loop quantum gravity (LQG) treatments \cite{9801080}-\cite{0905.3168}. \\

\noindent
In thermodynamics literature, one comes across
various statistical ensembles, such as the canonical, grand canonical, isothermal-isobaric and other distinct variants of the Gibbs ensemble~\cite{Hill1}-\cite{Hill2}.
A common thread of all the aforementioned ensembles is that there is at least one extensive variable which is exchanged
\footnote{Let us also note that in the microcanonical ensemble, none of the extensive variables are allowed to fluctuate. However, we do not consider it here.}
by the system with the surroundings. It could be energy, volume, number of particles or other conserved charges. Little attention has been paid to studying
thermodynamics of general systems and in particular black holes in an ensemble where the control parameters are intensive and independent, such as chemical potential $\mu$, pressure $P$ and temperature $T$. This is the so called open or $(\mu,P,T)$-ensemble. An open ensemble can be properly defined for small systems \cite{Hill2}, confined systems \cite{confined} or non-additive systems \cite{long,long1} including the case of black holes where entropy is not a homogeneous function of the energy, volume and other charges. The case of black holes is one example where the usual Gibbs-Duhem relation does not hold and consequently, the intensive parameters $\mu$, $P$ and $T$ are all independent. As we review below in the case of black holes, such systems have a non-trivial energy function called the subdivision potential~\cite{Hill2} or replica energy $\mathcal{E}$~\cite{Ruffo1,Ruffo2} in statistical mechanics. Conventional thermodynamics deals with the situation where $\mathcal{E}=TS$, ensuring that the associated free energy is trivial. However, for non-additive systems such as black holes, a non-zero free energy function can be obtained. \\

\noindent
In a preceding work, we studied the general form of logarithmic corrections to black hole entropy in the isothermal-isobaric ensemble where the system may exchange energy and thermodynamic volume with the environment which is parametrized by temperature $\beta$ and pressure $P$. Here, the thermodynamic pressure is provided by the cosmological constant \(\Lambda\) as \cite{Kastor:2009wy}-\cite{Kubiznak:2012wp},
\begin{equation}
P = - \frac{\Lambda}{8\pi}\,= \frac{(d-1)(d-2)}{16\pi l^2}.
\end{equation}
In order to obtain the perturbative entropy corrections for black holes, we adopt a rather macroscopic approach in which the knowledge of the thermodynamic free energy of the black hole shall allow us to determine the extent of thermodynamic fluctuations of the system about its equilibrium in terms of familiar second order response functions from classical thermodynamics (see \cite{parthasarathi}-\cite{Ghoshentropy} for some related works).
To leading order, the corrected entropy takes the general form,
\begin{equation}\label{zzz}
  \mathcal{S} = S_0 - k \ln S_0 + \cdots
\end{equation} where \(k\) is a constant which depends on the specific black hole under consideration and \(``\cdots"\) indicates other terms which might be present in
this order \cite{parthasarathi,Ghoshentropy}. It should be noted
that in eqn
(\ref{zzz}), \(S_0\) is the entropy of the black hole given by the Bekenstein-Hawking area rule, i.e. it is equal to a quarter of the area of the black hole horizon. \\

\noindent
The aim of the present work is to obtain the general form of logarithmic corrections to the black hole entropy function by considering fluctuations in energy, thermodynamic volume and other conserved charges in AdS black holes in the \((\mu,P,T)\)-ensemble where the \(\mu\) can be either the electric potential $\Phi$ (for charged black holes) or angular velocity $\Omega$ (for rotating black holes).\\

\noindent
The plan of this paper is as follows. In section-(\ref{mupt}), we set up the thermodynamics in the \((\mu,P,T)\)-ensemble by taking the system to be in contact with baths for energy, volume and number of particles/charges. The exact analytic expressions for the (co-)variances associated with fluctuations of energy, volume and number of particles or other conserved charges are computed. These are then used to find the general logarithmic corrections to the entropy of any thermodynamic system in the \((\mu,P,T)\)-ensemble. In the following section-(\ref{RNADS}), we first briefly discuss the thermodynamics of charged black holes in AdS and show the non-equivalence of the \((Q,P,T)\)- and \((\Phi,P,T)\)-ensembles where \(Q\) is the electric charge and \(\Phi\) is the electric potential of the black hole. All the second order response functions are found to diverge at the certain special points in the \((\Phi,P,T)\)-ensemble and the region of thermodynamic stability is identified. Next, by explicitly computing the elements of the covariance matrix in the \((\Phi,P,T)\)-ensemble, we find the exact expression of the log corrected black hole entropy for charged AdS black holes in four dimensions, which are valid in the stable region of the system. Subsequently in section-(\ref{rotating}), our analysis is generalized to rotating black holes with fluctuating angular momentum and we then compute the logarithmic entropy corrections for rotating BTZ and four dimensional Kerr-AdS black holes in the \( (\Omega,P,T)\)-ensemble in small angular momentum limit. We conclude with some discussions in section-(\ref{conclusions}). Calculation of the elements of the covariance matrix is summarized in appendix-(\ref{appendixa}). The Maxwell relations used in section-(\ref{mupt}) for simplifying the computations are collected in appendix-(\ref{appendixb}).

\section{Thermodynamic fluctuations and entropy corrections in the open ensemble} \label{mupt}
In this section, we compute the lowest order perturbative corrections to the entropy function of a thermodynamic system in the open ensemble. Such entropy corrections arise due to fluctuations in energy, thermodynamic volume and number of particles or the conserved charges in the \((\mu, P, T)\)-ensemble. It should be specially emphasized however, that such corrections are valid sufficiently far away from the critical point at which the response functions diverge signalling the onset of strong fluctuations. To start with, let us consider the system (which might be a black hole) in contact with energy, volume and particle (or say, electric charge/angular momentum for black holes) reservoirs with fixed intensive parameters respectively \(\beta\), \(P\) and \(\mu\). Then the most natural partition function for this \((\mu,P,T)\)-ensemble which we denote by the symbol \(\mathcal{Z} = \mathcal{Z}(\beta \mu, \beta P, \beta)\) is defined as (up to an overall constant factor to make \(\mathcal{Z}\) dimensionless),
\begin{equation}\label{abc}
  \mathcal{Z}(\beta \mu, \beta P, \beta) = \int \int Z(N,V,\beta) e^{-\beta(PV - N \mu)} dN dV
\end{equation} where \(Z(N,V,\beta)\) is the canonical partition function. Note that we are letting \(N\) vary continuously. We also note that the free energy in this ensemble is defined as,
\begin{equation}
  \mathcal{F} := - \frac{\ln \mathcal{Z}}{\beta} = E + PV - N \mu - TS.
\end{equation} Typically, the quantity \(E + PV - \mu N - TS = 0\) due to the Euler relation in standard thermodynamics. That is why this `open' ensemble or the \((\mu, P, T)\)-ensemble is not discussed in elementary treatments of statistical mechanics. As we see in the next section, this ensemble is perfectly regular for the case of black holes in AdS. The \((\mu, P, T)\)-partition function of eqn (\ref{abc}) can be written as,
\begin{equation}\label{PFOE}
  \mathcal{Z}(\beta \mu, \beta P, \beta) = \int \int \int \rho(N,V,E) e^{-\beta(E + PV - N \mu)} dE dV dN\, .
\end{equation}
We have used the fact that the canonical partition function has the following definition,
\begin{equation}
  Z(N,V,\beta) = \int \rho(E,V,N) e^{-\beta E} dE
\end{equation}
as the Laplace transform of the density of states. One can then invert eqn (\ref{PFOE}) to find the density of states as,
\begin{equation}\label{def}
  \rho(N,V,E) = \frac{1}{(2\pi i)^3} \int \int \int \mathcal{Z}(\beta \mu, \beta P, \beta) e^{\beta(E + PV - N \mu)} d\beta d(\beta P) d(\beta \mu).
\end{equation}
Now using the fact that \(S(\beta \mu, \beta P, \beta) = \ln \mathcal{Z}(\beta \mu, \beta P, \beta) + \beta (E + PV - N \mu)\), eqn (\ref{def}) becomes,
\begin{equation}\label{ghi}
  \rho(N,V,E) = \frac{1}{(2\pi i)^3} \int \int \int e^{S(\beta \mu, \beta P, \beta)} d\beta d(\beta P) d(\beta \mu).
\end{equation}
Next, we expand the entropy about its equilibrium value \(S_0\) as,
\begin{equation}
  S(\beta \mu, \beta P, \beta) \approx S_0 + \frac{\partial^2 S}{\partial x^i \partial x^j}\Bigg|_{x^i = (x^i)_0; x^j = (x^j)_0} (x^i - (x^i)_0) (x^j - (x^j)_0) + \cdots \,
\end{equation} where \(i,j = 1,2,3\) with \(x^1 = \beta, x^2 = \beta P\) and \(x^3 = \beta \mu\).  Substituting this into eqn (\ref{ghi}) and performing the integrals, we have,
\begin{equation}
  \rho(N,V,E) = \frac{e^{S_0}}{\sqrt{2 \pi D}}
\end{equation} where \(D\) is the determinant of the \(3 \times 3\) covariance matrix whose elements are defined as,
\begin{equation}
 D_{ij} = \frac{\partial^2 S}{\partial x^i \partial x^j} \Bigg|_{x^i = (x^i)_0; x^j = (x^j)_0}.
\end{equation} Having obtained the density of states, one can straightforwardly define the microcanonical entropy of the system as its logarithm, i.e.
\begin{equation}\label{entropycorrected}
  \mathcal{S} := \ln \rho(N,V,E) = S_0 - \frac{1}{2} \ln D + ({\rm higher~order~terms}).
\end{equation}
We should remark here that these corrections are obtained in the Gaussian approximation, i.e. fluctuations up to the lowest order are taken into account. These corrections shall therefore, severely break down close to the critical point where fluctuations are large or points at which second order response functions such as the specific heat blow up indicating peculiar thermodynamic behavior. In such situations, it may not be justified to consider the effect of only the lowest order. \\
\begin{table*}[t]
\caption{Exact expressions for  \(\langle (\Delta y_i)(\Delta y_j) \rangle \) where \(y_1 = E\), \(y_2 = V\) and \(y_3 =N\).}
\centering
\begin{tabular}{| c | c | c | c | c |}
\hline
&\\
{\(\langle (\Delta y_i)(\Delta y_j) \rangle \)}   & {Expression} \\  & \\ \hline
&\\
\(\langle (\Delta E)^2 \rangle\)    & \(T^2
   C_{P,\mu} - 2 P T^2 V \alpha_V + 2 \mu T^2 \bigg(\frac{\partial N}{\partial T}\bigg)_{P,\mu} + PT \bigg[ P V \kappa_{T,\mu} + \mu \bigg(\frac{\partial N}{\partial P}\bigg)_{T,\mu}\bigg]\) \\
   & \(+ \mu T \bigg[- P \bigg(\frac{\partial V}{\partial \mu}\bigg)_{T,P} + \mu \chi_{T,P}\bigg]\) \\ &\\
%&\\
\(\langle (\Delta V)^2 \rangle\)    & \(TV \kappa_{T, \mu}\)  \\  &\\
%&\\
\(\langle (\Delta N)^2 \rangle\)    &  \(T \chi_{T, P}\) \\ &\\
%&\\
\(\langle (\Delta E)(\Delta V) \rangle\)          &  \(V T^2 \alpha_V - PVT \kappa_{T,\mu} + \mu T \bigg(\frac{\partial V}{\partial \mu}\bigg)_{T,P}\)    \\ &\\
%&\\
\(\langle (\Delta E)(\Delta N) \rangle\)         &   \( -T^2 \bigg(\frac{\partial N}{\partial T}\bigg)_{P,\mu} + PT \bigg(\frac{\partial V}{\partial \mu}\bigg)_{T,P} - \mu T \chi_{T,P}\)   \\ &\\
%&\\
\(\langle (\Delta V)(\Delta N) \rangle\)  & \(-T \bigg(\frac{\partial V}{\partial \mu}\bigg)_{T,P}\)  \\ &\\ \hline
\end{tabular}\label{table1}
\end{table*}

Now in the present case, in order to compute exact expressions for the corrected microcanonical entropy with the corrections arising due to thermodynamic fluctuations, we need to compute the quantity \(D\) for various systems. To do this, let us first note that,
\begin{equation}
D_{ij} =  \frac{\partial^2 S}{\partial x^i \partial x^j}\Bigg|_{x^i = (x^i)_0; x^j = (x^j)_0} = \frac{\partial^2 \ln \mathcal{Z}}{\partial x^i \partial x^j}\Bigg|_{x^i = (x^i)_0; x^j = (x^j)_0}.
\end{equation}
Denoting the elements of the covariance matrix as \(D_{ij} = \langle (\Delta y_i)(\Delta y_j) \rangle \) where \(y_1 = E\), \(y_2 = V\) and \(y_3 = N\), their exact expressions are listed in table-(1) (see appendix-(\ref{appendixa}) for details). Once the elements of the covariance matrix are computed, they can subsequently be used to obtain the exact logarithmic correction to entropy of general thermodynamic systems. 
Here, \(\alpha_V = (1/V)(\partial V/\partial T)_{P,\mu}\) is the volume expansivity at fixed \(P\) and \(\mu\), \(C_{P,\mu} = T (\partial S/\partial T)_{P,\mu}\) is the specific
heat at fixed \(P\) and \(\mu\), \(\kappa_{T,\mu} = - (1/V) (\partial V/\partial P)_{T,\mu} \) is the compressibility at fixed
\(T\) and \(\mu\) and \(\chi_{T,P} = (\partial N/\partial \mu)_{T,P}\) is the susceptibility at fixed \(T\) and \(P\).\\

In the next two sections, we shall compute logarithmic corrections to the entropy of charged and rotating black holes due to thermodynamic fluctuations around their equilibrium using the techniques developed in this section. However, it should be noted that in the treatments to follow, black holes shall be treated as thermodynamic systems within the realm of classical equilibrium thermodynamics. While this is a good description at higher temperatures, as one approaches extremality, i.e. \(T \rightarrow 0\), strong quantum mechanical behavior becomes unavoidable \cite{extremal1,extremal2}. In such a regime, our results shall break down and one expects consistent results only from a quantum theory of gravity. Therefore, one should bear in mind that our results are valid quite far away from extremality. In what follows, it would be implicitly assumed that this is indeed the case.

\section{Charged black holes in the $(\Phi,P,T)$-ensemble}\label{RNADS}
Let us start by considering for example, the case of charged black holes in AdS, by making the identifications \(\mu \rightarrow \Phi\) and \(N \rightarrow Q\), with $\Phi$ denoting the electric potential. We note that the black hole mass is not a genuinely homogeneous function\footnote{It is rather quasi-homogeneous. See for example \cite{quasi} and references therein.} of the extensive variables but respects a slightly different relation known as the Smarr formula~\cite{Smarr} for the ADM mass which in \(d\)-dimensions reads \cite{Kastor:2009wy,Gunasekaran:2012dq},
\begin{equation}
  (d-3)M = (d-2)TS - 2PV + (d-3)Q \Phi.
\end{equation}
Consequently, it is possible for us to work in the \((\Phi, P, T)\)-ensemble for charged black holes with a non-trivial free energy function\footnote{More generally, the free energy can be computed from the Euclidean action \cite{York,Chamblin} (see also \cite{Kubiznak:2012wp}).}. 
For simplicity, we consider the four dimensional case. The metric and the gauge field solving the Einstein-Maxwell equations with a negative cosmological constant in the static coordinates are well known and are given by, 
\begin{equation}
	ds^2 = - f(r) dt^2 + f(r)^{-1} dr^2 + r^2 d\Omega_2 ^2, \hspace{5mm} {\rm and,} \hspace{4mm} F = dA
\end{equation} where \(d \Omega^2 = d\theta^2 + \sin ^2 \theta d\phi^2\) and, 
\begin{equation}
f(r) = 1 - \frac{2m}{r} + \frac{q^2}{r^2} + \frac{r^2}{l^2}, \hspace{8mm}  A = \frac{q}{r} dt.
\end{equation}
Here, the parameter \(m\) is the ADM mass of the spacetime, i.e. \(M = m\) and can be found by setting \(f(r_+) = 0\) where \(r_+\) is the largest root of \(f(r)\). In four dimensions, the metric parameter \(q\) exactly corresponds to the electric charge of the black hole, i.e. \(Q = q\).

\subsection{Thermodynamics and stability}
Within the framework of extended thermodynamics, the cosmological constant of the AdS spacetime is treated as a thermodynamic pressure and the ADM mass of the spacetime is appropriately identified with the enthalpy \(M := H = E + PV\) satisfying \(dM = TdS + VdP + \Phi dQ\). The natural candidate for energy function in this ensemble is therefore \(\mathcal{E} := M - Q \Phi = E + PV - Q \Phi\) which satisfies \(d\mathcal{E} = TdS + VdP - Q d\Phi\). Thus, the free energy in the \((\Phi, P, T)\)-ensemble for a charged AdS black hole takes the generic form,
\begin{equation}
  \mathcal{F} = M - Q \Phi - TS
\end{equation} and satisfies the first law,
\begin{equation}
  d\mathcal{F} = - SdT + VdP - Q d\Phi.
\end{equation} 
Therefore, the following response functions can be defined as its second derivatives,
\begin{equation}\label{response}
  C_{P,\Phi} = - T \bigg(\frac{\partial^2 \mathcal{F}}{\partial T^2}\bigg)_{P,\Phi}, \hspace{5mm} \kappa_{T, \Phi} = - \frac{1}{V} \bigg(\frac{\partial^2 \mathcal{F}}{\partial P^2}\bigg)_{T,\Phi}, \hspace{5mm} \chi_{T,P} = - \bigg(\frac{\partial^2 \mathcal{F}}{\partial \Phi^2}\bigg)_{T,P}.
\end{equation}
%%%%%%%%%%%%%%%%%%
Here, \(C_{P,\Phi}\) is the specific heat at constant pressure and electric potential whereas, \(\kappa_{T, \Phi}\) and \(\chi_{T,P}\) respectively are the isothermal-constant potential compressibility and susceptibility at constant temperature and pressure. Although, the above expressions are written with charged AdS black holes in mind, analogous ones can be obtained for other cases. 
%For example, the above analysis goes through for the case of Kerr-AdS black holes considered in subsection-(\ref{Kerr}) with the trivial identification $Q \rightarrow J$ and $\Phi \rightarrow \Omega$.
Using standard thermodynamic definitions, the expression for \(M\) expressed as a function of \(S\), \(P\) and \(Q\) reads~\cite{Chamblin,Chamblin:1999hg,Kubiznak:2012wp},
\begin{equation}\label{HRN}
M(S,P,Q)=\frac {1} {6\sqrt{\pi}} S^{-\frac{1}{2}}
\left(8 P S^2 + 3S+3\pi Q^2 \right) \,
\end{equation}
where one uses,
\begin{equation}
  S = \pi r_+^2 \hspace{5mm} {\rm and,} \hspace{4mm} P = \frac{3}{8 \pi l^2}.
\end{equation} Thus, the Hawking temperature \(T\), thermodynamic volume \(V\) and electric potential \(\Phi\) are defined by taking partial derivatives of \(M\) with respect to \(S\), \(P\) and \(Q\) respectively. They are given by,
\begin{equation} \label{tQ}
 T =  \frac{1}{4\sqrt{\pi}} \, S^{-3/2} \left( 8 P S^2 + S - \pi Q^2 \right), \hspace{4mm}  V= \frac{4 S^{3/2}}{3 \sqrt{\pi }}, \hspace{4mm} \Phi = \frac{\sqrt{\pi} Q}{\sqrt{S}}.
\end{equation}
The energy function \(\mathcal{E}\) of the \((\Phi, P, T)\)-ensemble can be obtained by the Legendre transform \(\mathcal{E} = M - Q \Phi\) thus giving,
\begin{equation}
  \mathcal{E}(S,P,\Phi) = \frac {1} {6\sqrt{\pi}} S^{-\frac{1}{2}}
\left(8 P S^2 + 3S-3 S \Phi^2 \right).
\end{equation}
\noindent
It should specially be remarked that for the case of charged black holes in AdS spacetimes, due to the dependence of the thermodynamic volume \(V\) on the entropy \(S\), the specific heat(s) at constant \(V\) are trivial, i.e.
\begin{equation}
  C_{V,Q} = \bigg(\frac{\partial E}{\partial T}\bigg)_{V,Q} = T \bigg(\frac{\partial S}{\partial T}\bigg)_{V,Q} = 0
\end{equation} and
\begin{equation}
  C_{V,\Phi} = \bigg(\frac{\partial U}{\partial T}\bigg)_{V,\Phi} = T \bigg(\frac{\partial S}{\partial T}\bigg)_{V,\Phi} = 0.
\end{equation}
%As was also done in [ ], for the sake of regularizing our results, we shall take the constant volume specific heat to be an infinitesimally small positive number, say \(\epsilon\) which is arbitrarily close to zero yet is not identically equal to zero.
We also have the following specific heats,
\begin{equation}
  C_{P,Q} = \bigg(\frac{\partial M}{\partial T}\bigg)_{P,Q} = T \bigg(\frac{\partial S}{\partial T}\bigg)_{P,Q}, \hspace{5mm} C_{P,\Phi} = \bigg(\frac{\partial \mathcal{E}}{\partial T}\bigg)_{P,\Phi} = T \bigg(\frac{\partial S}{\partial T}\bigg)_{P,\Phi}
\end{equation}
%With thermodynamic variables \(\{T,S,P,V,\Phi,Q\}\), one can derive several Maxwell-like relations which are listed in the appendix.
which are calculated to be,
\begin{equation}
C_{P,Q}=C_{V,Q}+\frac{2 S \left(8 P S^2-\pi  Q^2+S\right)}{S (8 P S-1)+3 \pi  Q^2}
\end{equation} and,
\begin{equation}
C_{P,\Phi}=C_{V,\Phi}+\frac{2 S \left(8 P S-\Phi ^2+1\right)}{8 P S+\Phi ^2-1}.
\end{equation}
Here one has \(C_{V,Q} = C_{V,\Phi} = 0\) but we include them for later use. 
%They are plotted in figure-(\ref{CpQphi_smal}) and show similar qualitative behavior. 
It is to be particularly noted that the specific heats \(C_{P,Q}\) and \(C_{P,\Phi}\) diverge at different points. This is because the \((Q,P,T)\)- and \((\Phi,P,T)\)-ensembles have distinct physical behavior. The non-equivalence of the fixed charge \((Q,T)\)- and fixed potential \((\Phi,T)\)-ensembles for charged AdS black holes is well known \cite{Chamblin,Caldarelli:1999xj} (see also \cite{Caicharged}), and these features carry over to the \((Q,P,T)\)- and \((\Phi,P,T)\)-ensembles being studied here. 
%The specific heat \(C_{P,Q}\) diverges at the critical point of the \((Q,P,T)\)-ensemble which is useful to note. 
Our main interest in this work is in the phase structure in the $(\Phi,P,T)$ ensemble, but few comments are also made in the fixed charge ensemble below for completeness and to put our results in perspective.\\

\begin{figure}[h!]
	% \begin{wrapfigure}{r}{0.43\textwidth}
	%\begin{center}
	{\centering
		\subfloat[]{\includegraphics[width=4in]{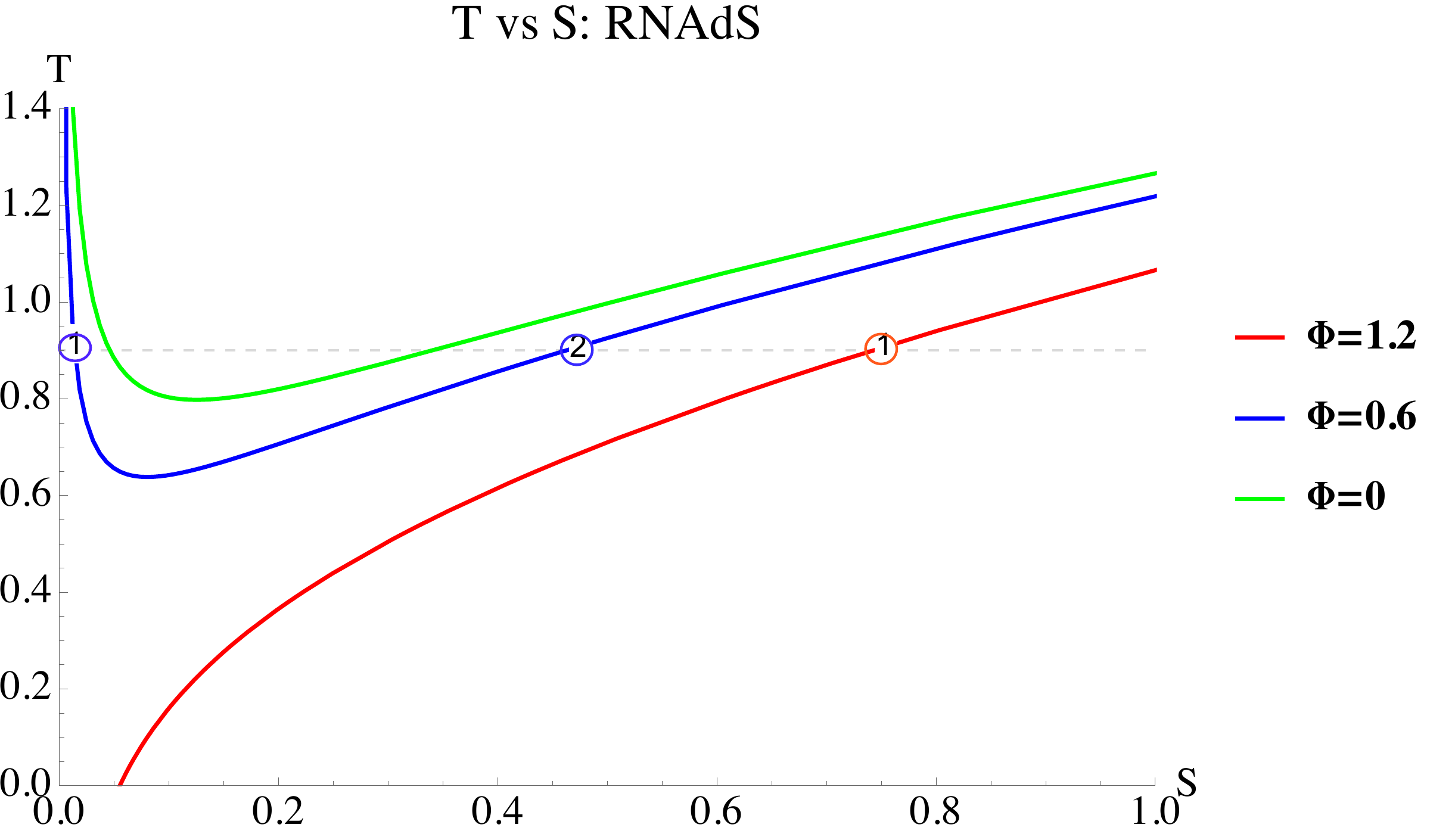}\label{t_Phi}}\hspace{0.5cm}	
		\subfloat[]{\includegraphics[width=4in]{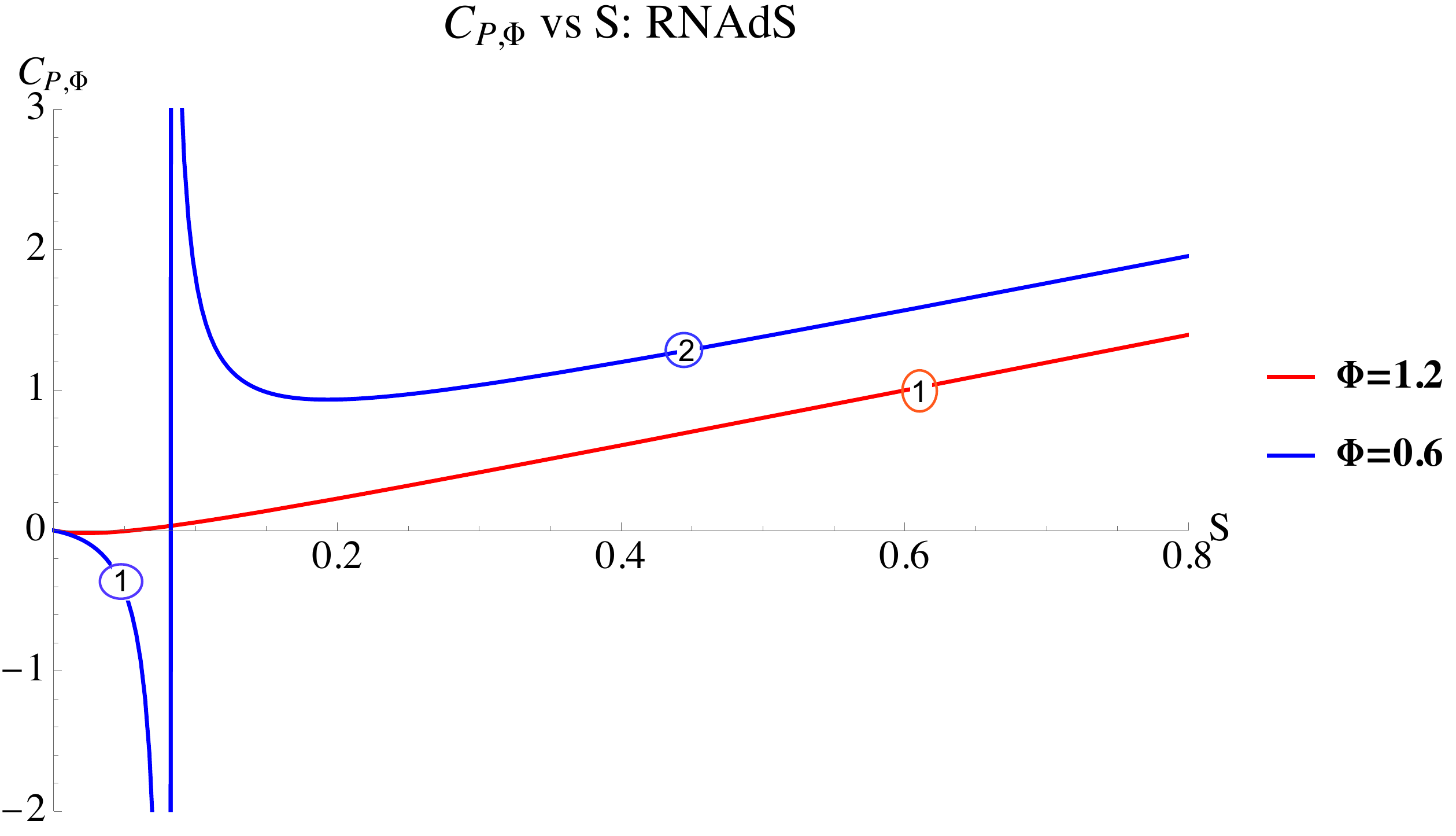}\label{Cp_Phi}}				
	% 2a,2b	
		
		\caption{\footnotesize (a) Temperature vs entropy for charged black holes in AdS for $P=1$ and various values of potential $\Phi$. The thermodynamics of the system is distinctly different depending on whether $\Phi <1$ or $\Phi >1$. For $\Phi <1$ and $T=0.82$, the smaller and larger black holes have entropy $S=0.018$ and $S=0.35$, respectively. (b) The specific heat \(C_{P,\Phi}\) is plotted versus \(S\) for $\Phi < 1$ and $\Phi > 1$, from which the thermodynamic stability of the solutions indicated in figure-(\ref{t_Phi}) can be understood. \(C_{P,\Phi}\) does not diverge anywhere for $\Phi > 1$, whereas it diverges at $S=0.08$ for $\Phi < 1$. This value of entropy corresponds to the equilibrium points in figure-(\ref{t_Phi}), where new black hole phases appear.
	}}
\end{figure}

\noindent
{\underline {Phase structure}}:   Figures-(\ref{t_Q}) and (\ref{t_Phi}) show the temperature as a function of entropy (horizon radius) for different values of charge and potential, respectively. In the fixed charge ensemble, there exists a critical point and phase structure is very different. The value of \(Q_{cr}\)  in figure-(\ref{t_Q}) can be determined straightforwardly from the respective conditions \cite{Chamblin}:
%%%%%%%%%%%%%%
\begin{figure}[h!]
	% \begin{wrapfigure}{r}{0.43\textwidth}
	%\begin{center}
	{\centering
		\subfloat[]{\includegraphics[width=4.0in]{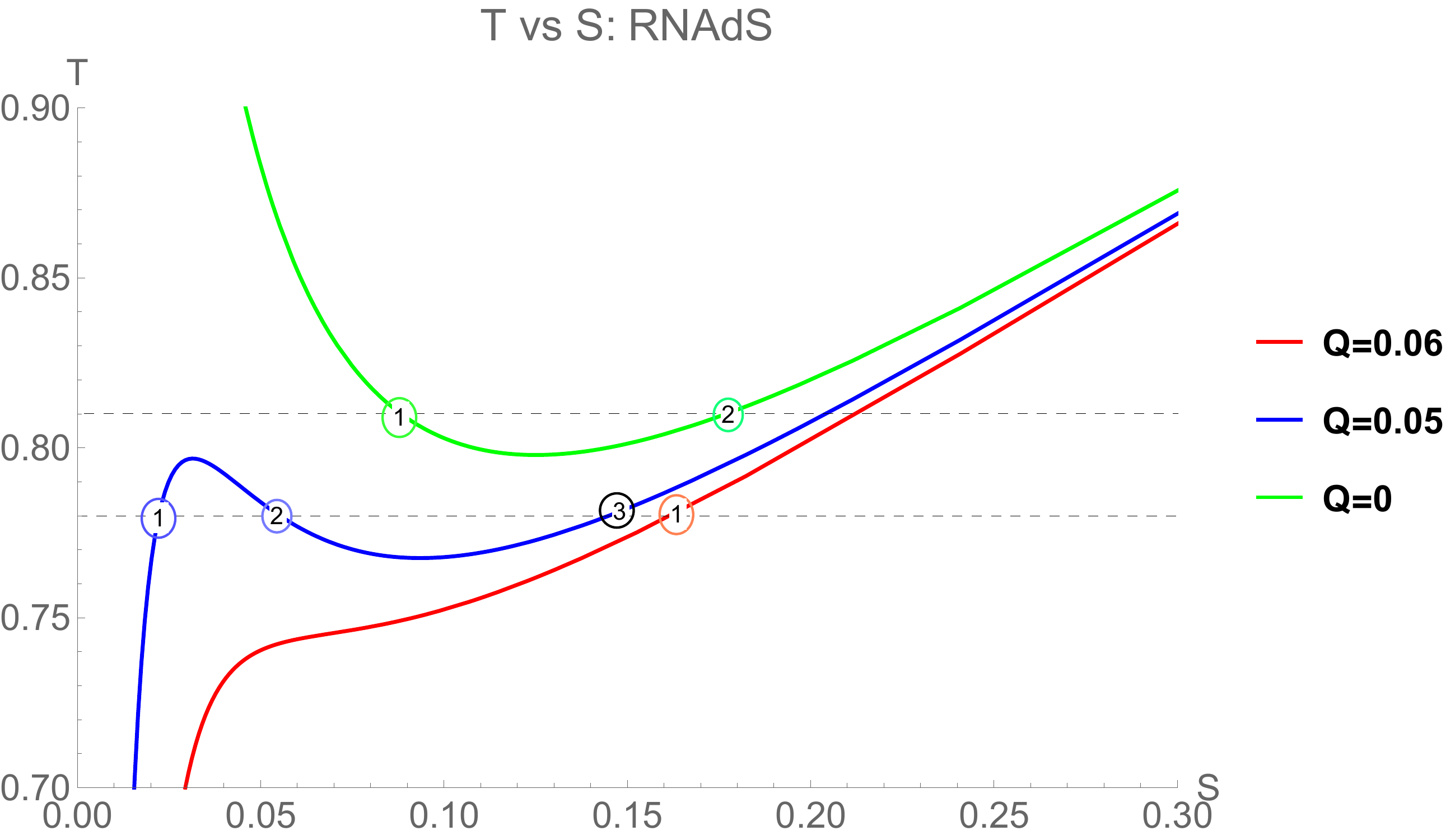}\label{t_Q}}\hspace{0.5cm}	
		\subfloat[]{\includegraphics[width=4.0in]{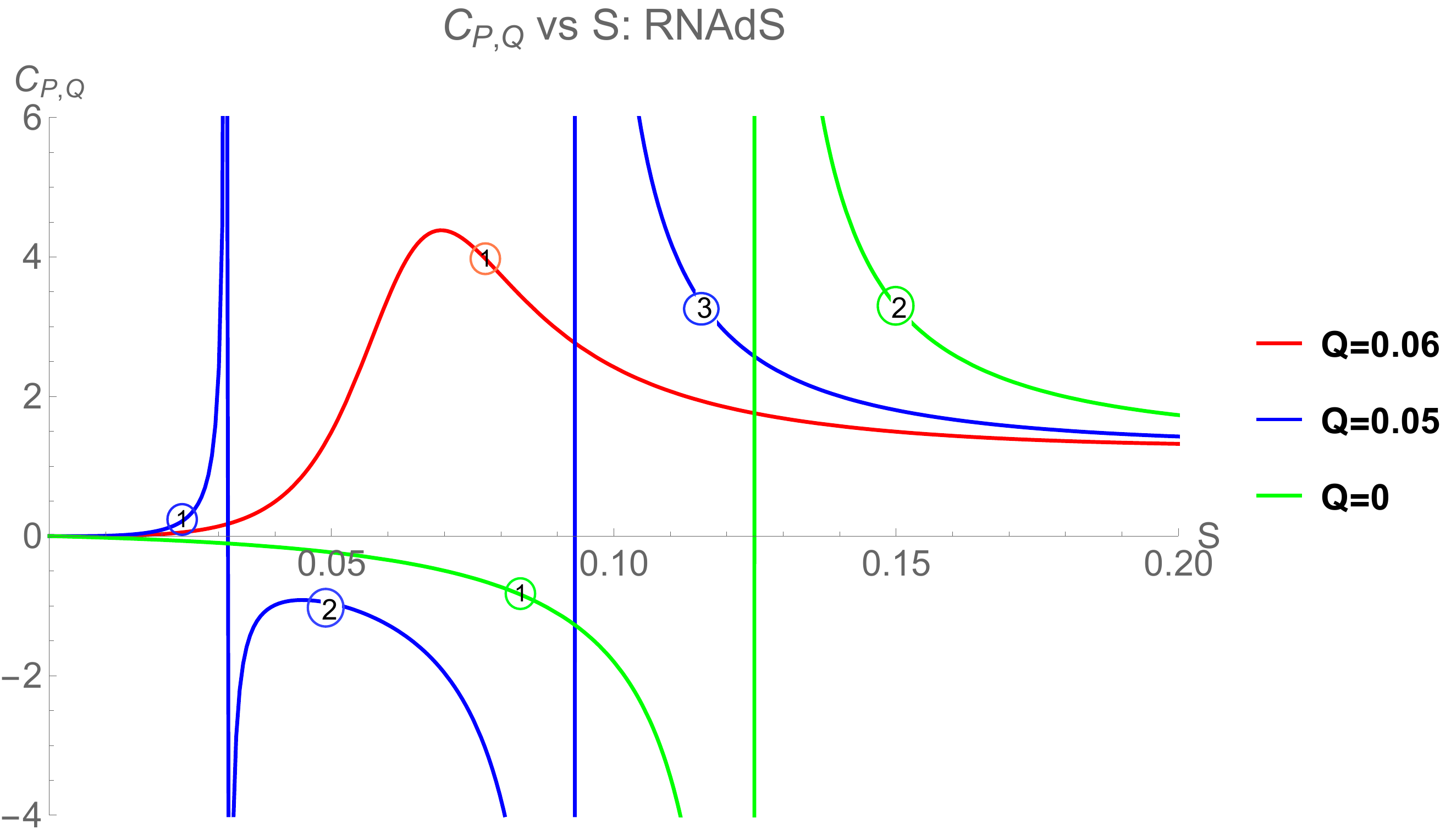}\label{Cp_Q}}				
	% 2a,2b	
		
		\caption{\footnotesize (a) Temperature vs entropy for charged black holes in AdS for $P=1$ and various charge parameters greater and less than $Q_{cr}$. The value of critical charge for above choice of parameters is $Q_{cr}=0.0575824$. At a given temperature, depending on the value of charge, there can be up to three branches of black hole solutions. For $Q=0$ case and $T=0.82$ the small and large black hole branches appear at $S=0.078$ and $S=0.199$ respectively. For $Q=0.05$ and $T=0.78$, there is a small branch at $S=0.02$, an intermediate branch at $S=0.05$ and a large branch at $S=0.14$. For $Q=0.06$ and any temperature, there is always a single large black hole branch. (b) The specific heat \(C_{P,Q}\) is plotted versus \(S\) for $Q=0$, $Q < Q_{cr}$ and $Q> Q_{cr}$ showing the stability of solutions. \(C_{P,Q}\) diverges at two points, namely $S=0.03$ and $S=0.9$, which are equilibrium points indicating new phases of the system. When $Q=Q_{cr}$, \(C_{P,Q}\) diverges at a single point $S=0.06$ (not shown in the plot). } 
	}
\end{figure}
\begin{equation} \label{dTdS}
  \bigg(\frac{\partial T}{\partial S}\bigg)_{Q,P}= \bigg(\frac{\partial^2 T}{\partial S^2}\bigg)_{Q,P} = 0 \, ,
\end{equation}
and turns out to be $Q_{cr}=\frac{1}{4 \sqrt{6 \pi } \sqrt{P}}$. As can be seen from figure-(\ref{t_Q}), for $Q=0$, there are two branches, for $Q<Q_{cr}$, there are three branches and for $Q>Q_{cr}$ there is a single branch of black hole solutions. 
For instance for $Q=0$, there are the familiar small and large black hole branches, which are marked in the green plot in figure-(\ref{t_Q}).  For $Q=0$ case and $T=0.82$ the small and large black hole branches appear at $S=0.07$ and $S=0.19$, where the former is unstable and the later is stable, respectively. For $Q=0.05$ and $T=0.78$, there is a small branch at $S=0.02$, an intermediate branch at $S=0.05$ and a large branch at $S=0.14$. The small and large black hole branches are locally stable, whereas the intermediate branch is unstable. For $Q=0.06$ and any temperature, there is always a single large black hole branch. Stability of all the aforementioned branches of solutions can be seen clearly from figure-(\ref{Cp_Q}). \(C_{P,Q}\) diverges at two points, namely $S=0.03$ and $S=0.9$, which are equilibrium points indicating new phases of the system. At the critical value of charge $Q=Q_{cr}$, the two points merge and \(C_{P,Q}\) diverges at a single point (not shown in the plots).\\
%%%%%%%%%%%%%%
%%%%%%%%%%%%%%%%%%%%%%%%%%

\noindent
In the $(\Phi,P,T)$ ensemble, the temperature vs the entropy plot given in figure-(\ref{t_Phi}) reveals that there are essentially two regimes of potential, namely small and large. For the parameter choices made here, these are $\Phi <1$ and $\Phi >1$, respectively. The thermodynamics in the $\Phi <1$ is identical to the case with $\Phi =0$, i.e., there is a branch with a smaller radius and one with larger radius. The branch with smaller radius has negative specific heat and hence is unstable. The situation is similar to the case of Schwarzschild-AdS black holes where only above a certain transition temperature the free energy is negative and the stable large black hole branch dominates thermodynamics. This can be inferred from the plot of specific heat in figure-(\ref{Cp_Phi}) as well. For $\Phi >1$, there is however a unique stable black hole solution with larger radius, and this dominates the thermodynamics at all temperatures~\cite{Chamblin}. In this large potential regime, there are further two sub cases to consider: namely high and low temperatures. As can be seen from figures-(\ref{t_Phi}) and  (\ref{Cp_Phi}) (red curves), high temperatures aid stability and the non-extremal black holes dominate the physics. At low temperatures, as temperature goes to zero the entropy is non-zero (red curve in figure-(\ref{t_Phi})) and this indicates that the thermodynamic behavior is still dominated by a black hole, which is the extremal hole with finite radius. Setting $\Phi=1$ gives back the anti de-Sitter (AdS) spacetime. In this paper, we are interested in studying entropy corrections coming from thermodynamic as well as volume fluctuations for non-extremal black holes away from critical region, as they dominate the physics.\\

%%%%%%%%%%%%%%%%%%
%%%%%%%%%%%%%%
\begin{figure}[h!]
	% \begin{wrapfigure}{r}{0.43\textwidth}
	%\begin{center}
	{\centering
		\subfloat[]{\includegraphics[width=3in]{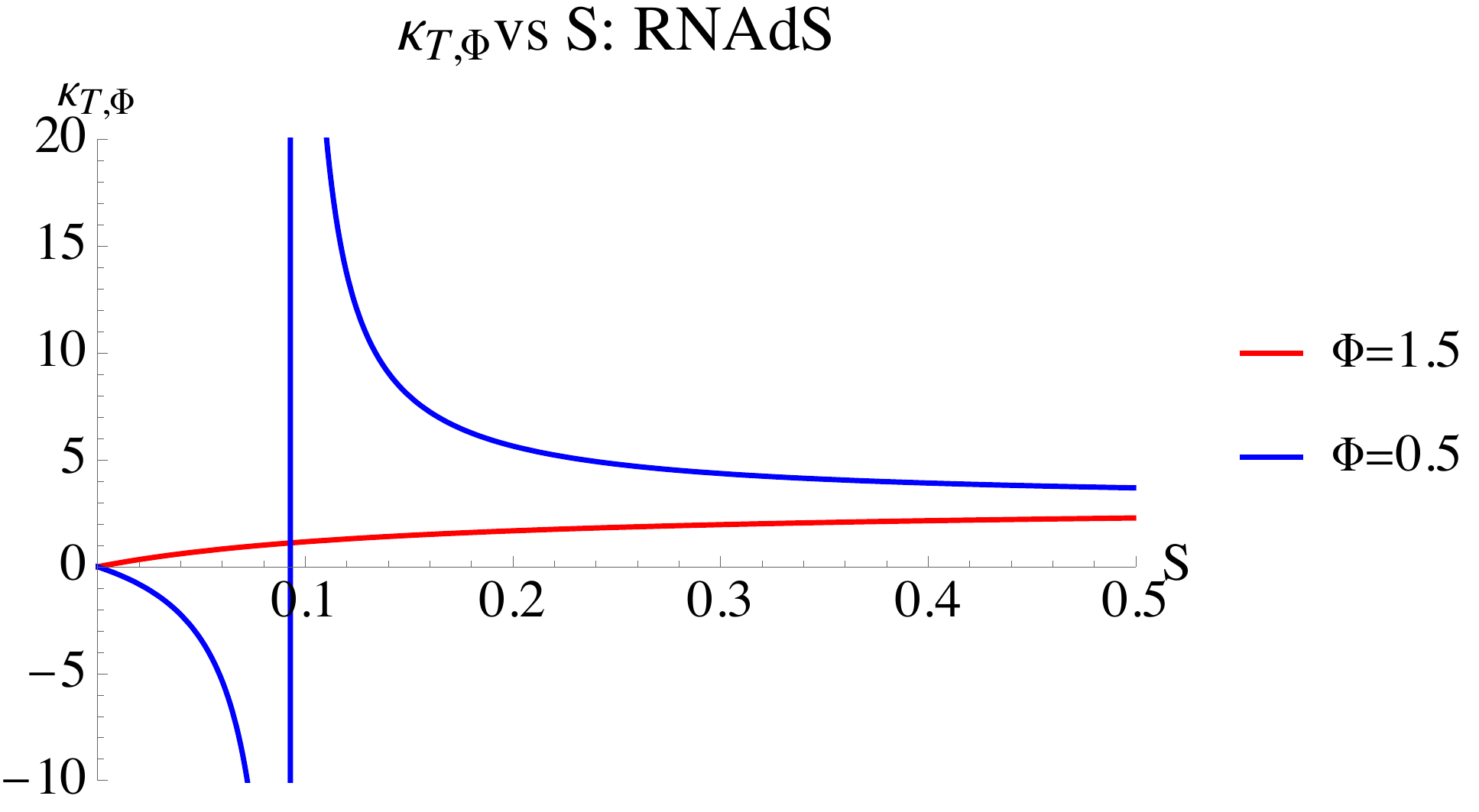}\label{Kappa_Phi}}\hspace{0.5cm}	
		\subfloat[]{\includegraphics[width=3in]{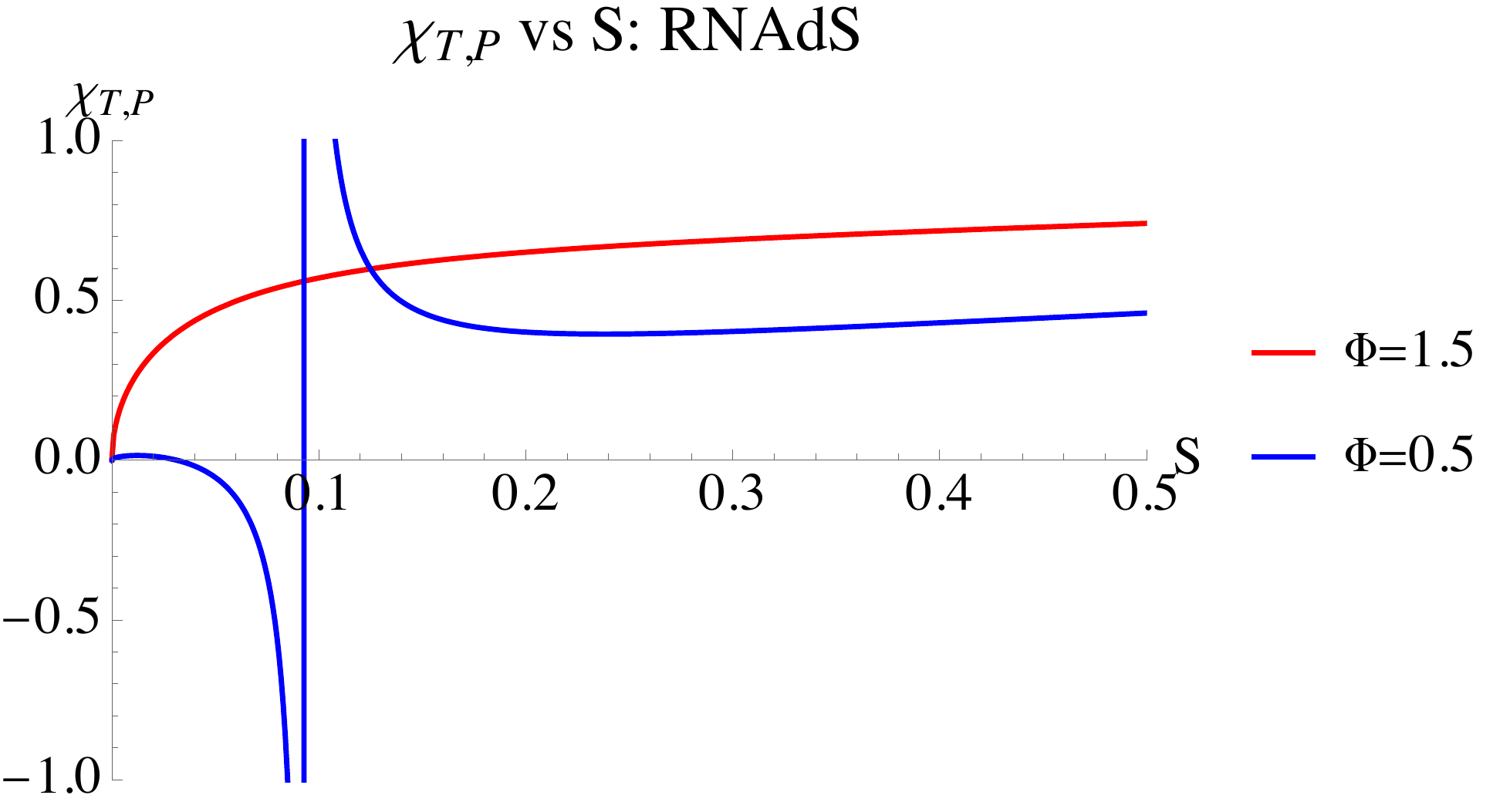}\label{Chi_TP}}				
	% 2a,2b	
		
		\caption{\footnotesize (a) Compressibility \(\kappa_{T,\Phi}\) for $\Phi<1$ and $\Phi>1$. (b) Susceptibility \(\chi_{T,P}\) for $\Phi<1$ and $\Phi>1$.} 
	}
\end{figure}
%%%%%%%%%%%%%%%%%%%%%%%%%%
\noindent
In addition to the specific heats, one has the response functions \(\kappa_{T, \Phi}\) and \(\chi_{T, P}\) whose exact expressions are,
\begin{equation}
\kappa_{\Phi,T}=\frac{24 S}{8 P S+\Phi ^2-1} \hspace{5mm} {\rm and,} \hspace{4mm} \chi_{T,P}=\frac{\sqrt{S} \left(8 P S+3 \Phi ^2-1\right)}{\sqrt{\pi } \left(8 P
   S+\Phi ^2-1\right)}.
\end{equation}
They have been plotted for various potential ranges in figures-(\ref{Kappa_Phi}) and (\ref{Chi_TP}). They too diverge at the same point as \(C_{P,\Phi}\). 
%Thus exactly at $\Phi=1$ in the fixed potential ensemble, all the second order response functions \(C_{P,\Phi}\), \(\kappa_{T,\Phi}\) and \(\chi_{T,P}\) are found to diverge. 
%This point greatly resembles the one pointed out by Davies \cite{Davies} for rotating black holes where the thermodynamically stable and unstable regions are separated by the critical point. This can clearly be seen from the behaviour of the response functions as a function of \(S\) (or \(r_+\)) with some of them being negative on one side of the critical point versus all of them being positive definite on the other side. 
It should be recalled that thermodynamic stability is ensured by the positivity of all the second order response functions arising in the ensemble of interest. In the present case, thermodynamic stability corresponds to the positivity of \(C_{P,\Phi}\), \(\kappa_{T,\Phi}\) and \(\chi_{T,P}\). They are all positive in the large black hole branch as seen in figures-(\ref{Cp_Phi}), (\ref{Kappa_Phi}) and (\ref{Chi_TP}), indicating towards the thermodynamic stability of the large black hole. On the other hand, the small black hole branch is thermodynamically unstable with negative values of the second order response functions. Similar conclusions can be made for the Kerr-AdS black hole discussed later~\cite{HawkingPage}. 
%However, one should remember that although the large black hole branch is always thermodynamically stable against fluctuations, it is not always the thermodynamically preferred phase (see for example \cite{HawkingPage}). The latter is thermal AdS for smaller sizes of the large black hole. \\

\subsection{Entropy corrections}
Now that we have exact expressions for all the response functions and other thermodynamic quantities for the four dimensional charged AdS black hole, it is straightforward to compute the individual fluctuations listed in table-(1) and hence obtain \(D\) exactly.  Let us note that although one has \(C_{V,\Phi} = 0\), for the sake of obtaining finite log corrections to the entropy function, we will adopt the strategy of taking \(C_{V,\Phi} = \epsilon\) where \(\epsilon\) is an infinitesimally small positive constant whose exact value is not important (see \cite{Wei1,Ghoshentropy}). Below, we give the final expression for \(D\) computed from the fluctuations of various quantities listed in table-(1), which reads,
\begin{equation}
D = \frac{S \epsilon \left(-8 P S+\Phi ^2-1\right)^4}{8 \pi ^3 \left(8 P S+\Phi
   ^2-1\right)}.
\end{equation}
It should be noted that \(D\) blows up at \(8 P S+\Phi^2-1 = 0\) which is equivalent to \((\partial T/\partial S)_{P,\Phi} = 0\). Consequently, these entropy corrections severely break down at this point. In fact, as also noted earlier, such perturbative corrections to the entropy can be expected to give sensible results only when one is sufficiently far away from such points where fluctuations around equilibrium are small and it suffices to consider only the effect of the lowest order. Therefore, our general method cannot be relied upon close to such points. Substituting the expression for \(D\) into eqn (\ref{entropycorrected}) and neglecting all constant terms, one finds,
\begin{equation}
  \mathcal{S} = S_0 - \frac{1}{2} \ln S_0 - 2 \ln \bigg[ -8 P S_0+\Phi ^2-1 \bigg] + \frac{1}{2} \ln \bigg[8 P S_0+\Phi
   ^2-1\bigg] + \cdots .
\end{equation}
Thus, the leading logarithmic contribution has a term of the form \(-k \ln S_0\) with \(k = 1/2\). It should be noted that in the above formula \(S_0\) is the entropy of the black hole at equilibrium and is given by the Bekenstein-Hawking formula. 

\section{Rotating black holes in AdS in the $(\Omega, P, T)$-ensemble} \label{rotating}
So far we have considered charged black holes in AdS and their thermodynamics in the \((\Phi,P,T)\)-ensemble where \(\Phi\) is the electric potential. It is straightforward to generalize such an analysis to rotating black holes with electric potential \(\Phi\) being replaced by the angular velocity \(\Omega\) of the rotating black hole \cite{Caldarelli:1999xj}. Recall that the mass \(M\) of the black hole satisfies \(dM = TdS + VdP + \Omega dJ\). The energy function \(\mathcal{E}\) of the \((\Omega,P,T)\)-ensemble is then defined as, \(\mathcal{E} = M - J \Omega\) and the free energy as \(\mathcal{F} = \mathcal{E} - TS = M - J \Omega - TS\). Thus, working in a similar fashion as before one can obtain the exact logarithmic corrections to the black hole entropy for rotating black holes in the open ensemble. Below, we separately consider the cases of the rotating BTZ black hole and the Kerr-AdS black hole in four dimensions.

\subsection{BTZ black holes}
We begin by recalling that BTZ black holes are solutions to (2+1)-dimensional Einstein gravity with a negative cosmological constant \cite{Banados:1992wn}. The metric reads, 
\begin{equation}
ds^2 = - f(r) dt^2 + f(r)^{-1} dr^2 + r^2 \bigg( d\phi - \frac{j}{2r^2} dt \bigg)^2
\end{equation} with metric function, 
\begin{equation}
f(r) = - 2m +\frac{r^2}{l^2} + \frac{j^2}{4r^2}.
\end{equation} Here, the parameters \(m\) and \(j\) map exactly to the mass and the angular momentum of the black hole, i.e. \(M = m\) and \(J = j\). The expression for the black hole mass as a function of \(S\), \(P\) and \(J\) is \cite{Dolan:2011xt,Dolan119,BTZ},
\begin{equation}
M(S,P,J) = \frac{\pi ^2 J^2}{128 S^2}+\frac{4 P S^2}{\pi }
\end{equation} where \(S = \pi r_+/2\). Thus,
\begin{equation}
T= \frac{8 P S}{\pi }-\frac{\pi ^2 J^2}{64 S^3}, \hspace{5mm} \Omega = \frac{\pi ^2 J}{64 S^2}, \hspace{5mm} V=\frac{4 S^2}{\pi }.
\end{equation}
The energy function \(\mathcal{E} = \mathcal{E}(S,P,\Omega)\) is therefore obtained to be,
\begin{equation}
  \mathcal{E}(S,P,\Omega) = \frac{4PS^2}{\pi} - \frac{32 S^2 \Omega^2}{\pi^2}.
\end{equation}
Consequently, in \((S,P,\Omega)\) variables, temperature and volume have expressions,
\begin{equation}
T=\frac{8 S \left(\pi  P-8 \Omega ^2\right)}{\pi ^2} \hspace{5mm} {\rm and,} \hspace{5mm} V = \frac{\pi ^3 T^2}{16 \left(\pi  P-8 \Omega ^2\right)^2}.
\end{equation}
The specific heat \(C_{P,\Omega}\) reads,
\begin{equation}
C_{P,\Omega}= \frac{\pi ^2 T}{8 \pi  P-64 \Omega ^2} + \epsilon
\end{equation} where we have put \(C_{V,\Omega} = \epsilon\) as before. The other response functions can similarly be calculated to give,
\begin{equation}
\kappa_{T,\Omega}= \frac{2 \pi }{\pi  P-8 \Omega ^2} \hspace{6mm} {\rm and,} \hspace{5mm} \alpha_V=\frac{2}{T}.
\end{equation}
Interestingly, the BTZ black hole does not admit any phase transition and therefore lacks the existence of a critical point \cite{BTZ}. However, note that both \(\kappa_{T,\Omega}\) and \(\alpha_V\) diverge at extremality but are positive definite otherwise. We can obtain the final expression for \(D\) to be (apart from some constants),
\begin{equation}
D = \frac{\epsilon  S^8 \left(\pi  P-8 \Omega ^2\right)^3}{\pi ^{10}}.
\end{equation}
Therefore, the log corrected entropy after neglecting constants and higher order terms is,
\begin{equation}
  \mathcal{S} = S_0 - 4 \ln S_0 + ({\rm terms~independent~of~} S_0).
\end{equation} Note that \(S_0\) is given by the Bekenstein-Hawking formula and we have suppressed the \(P\) and \(\Omega\) dependence. 

\subsection{Kerr-AdS black holes}\label{Kerr}
We now consider Kerr-AdS black holes in four dimensions and later on take the small angular momentum limit to get exact analytic results. In the Newman-Penrose formalism, the associated metric takes the following form in the Boyer-Lindquist coordinates~\cite{Caldarelli:1999xj,Gunasekaran:2012dq},
%%%%%%%%% EW
\begin{eqnarray}
ds^2&=&-\frac{\Delta}{\rho^2}(dt-\frac{a}{\Xi}\sin^2\!\theta d\varphi)^2+
\frac{\rho^2}{\Delta}dr^2+\frac{\rho^2}{\Sigma}d\theta^2 \nonumber \\
&+&
\frac{\Sigma \sin^2\!\theta}{\rho^2}[adt-\frac{(r^2+a^2)}{\Xi}d\varphi]^2\,.
\end{eqnarray}
Various quantities appearing in the metric are,
\begin{eqnarray}
\Delta&=&(r^2+a^2)(1+\frac{r^2}{l^2})-2mr\,,\quad
\Sigma=1-\frac{a^2}{l^2}\cos^2\theta \,,\nonumber \\
\Xi&=&1-\frac{a^2}{l^2}\,, \quad
\rho^2=r^2+a^2\cos^2\theta \,. 
\end{eqnarray}
\noindent
Here $l$ is the AdS length scalar, which is related to pressure $P=3/{8\pi l^2}$.
The thermodynamic quantities of the system are~\cite{Gunasekaran:2012dq,Altamirano:2014tva},
\begin{eqnarray}
M&=&\frac{m}{\Xi^2}\,, \quad J=\frac{ma}{\Xi^2} \,, \quad  \Omega=\frac{a}{l^2}\frac{r_+^2+l^2}{r_+^2+a^2}\,, \label{OHJ}\\
T&=&\frac{1}{2\pi r_+}\Bigr[\frac{(a^2+3r_+^2)(r_+^2/l^2+1)}{2(a^2+r_+^2)}-1\Bigr]\,, \label{BHT} \\
S &=&\pi \frac{(a^2+r_+^2)}{\Xi}=\frac{A}{4} \,,\quad V=\frac{r_+A}{3}\Bigl(1+\frac{1+r_+^2/l^2}{2r_+^2}\frac{a^2}{\Xi}\Bigr)\,. \label{BHS}
%G &=& \frac{(r^2_+ + 3a^2) l^4 -(r^2_+ - a^2)^2 l^2 + (a^2 + 3 r^2_+)a^2 r^2_+}{l^4 \Xi^2 r_+}\,, \nonumber \\
\end{eqnarray}
\noindent
Here $a$ is the rotation parameter characterizing the ergo-sphere of the rotating black hole and \(r_+\) is the horizon radius.
%\ba
%C_P=-\frac{2\pi l^4(a^2+r_+^2)^2[(r_+^2-l^2)a^2+r_+^2(l^2+3r_+^2)]}
%{(a^2-l^2)[(3a^4+6r_+^2a^2-r_+^4)l^4+(a^6+13r_+^2a^4+23r_+^4a^2+3r_+^6)l^2
%+a^2r_+^2(a^2+3r_+^2)^2]}\,.
%\ea
The thermodynamic behavior in the fixed $(J,P,T)$ ensemble is similar to the analysis presented for charged black holes in AdS in section-(\ref{RNADS}) with $J$ replacing $Q$. The case of slow rotation parameter $a << l$ or small angular momentum limit is quite interesting, and it has been shown earlier that the equation of state in this case reduces to \cite{Gunasekaran:2012dq}
\begin{eqnarray}
P&=&\frac{T}{v}-\frac{1}{2\pi v^2}+\frac{48J^2}{\pi v^6} \, 
\end{eqnarray}
where 
$v=2\Bigl(\frac{3V}{4\pi}\Bigr)^{\!1/3}$. For a fixed value of $J$, there is a critical point characterized by $(P_c, V_c, T_c)$, which can be determined. Further, for $P<P_c$ the system shows small black hole to large black hole phase transition akin to the liquid-gas transition. In this paper we do not pursue the cases of black holes which are either extremal or very close to extremality. In particular, in the latter case, it is known that the Kerr AdS black holes are unstable due to super radiant instabilities, with new stable configurations expected \cite{Hawking:1999dp,Sonner:2009fk,Dias:2010ma,Cardoso:2013pza}.
The black hole mass expressed as a function of entropy \(S\), pressure \(P\) and angular momentum \(J\) can be computed to be\cite{Dolan:2011xt,Dolan119,Gunasekaran:2012dq},
\begin{equation}
M(S,P,J) = \frac{\sqrt{(8 P S+3) \left(12 \pi ^2 J^2+S^2 (8 P
   S+3)\right)}}{6 \sqrt{\pi } \sqrt{S}}\,
\end{equation}
which in the small angular momentum limit can be truncated to,
\begin{equation}
M(S,P,J) = \frac{6 \pi ^2 J^2+S^2 (8 P S+3)}{6 \sqrt{\pi } S^{3/2}}
\end{equation}
from which one gets the angular velocity as $\Omega = \frac{2 \pi ^{3/2} J}{S^{3/2}}$ and thermodynamic volume as $V = \frac{4 S^{3/2}}{3 \sqrt{\pi }}$. Therefore, the energy function \(\mathcal{E}\) of the \((\Omega, P, T)\)-ensemble can be obtained by the Legendre transform \(\mathcal{E} = M - J \Omega\) thus giving,
\begin{equation}
 \mathcal{E}(S,P,\Omega)=\frac{\sqrt{S} \left(2 \pi  (8 P S+3)-3 S \Omega ^2\right)}{12 \pi ^{3/2}}.
\end{equation}
The Hawking temperature is,
\begin{equation}
T=\frac{2 \pi  (8 P S+1)-3 S \Omega ^2}{8 \pi ^{3/2} \sqrt{S}}
\end{equation}
\begin{figure}[t]
\begin{center}
\includegraphics[width=4.8in]{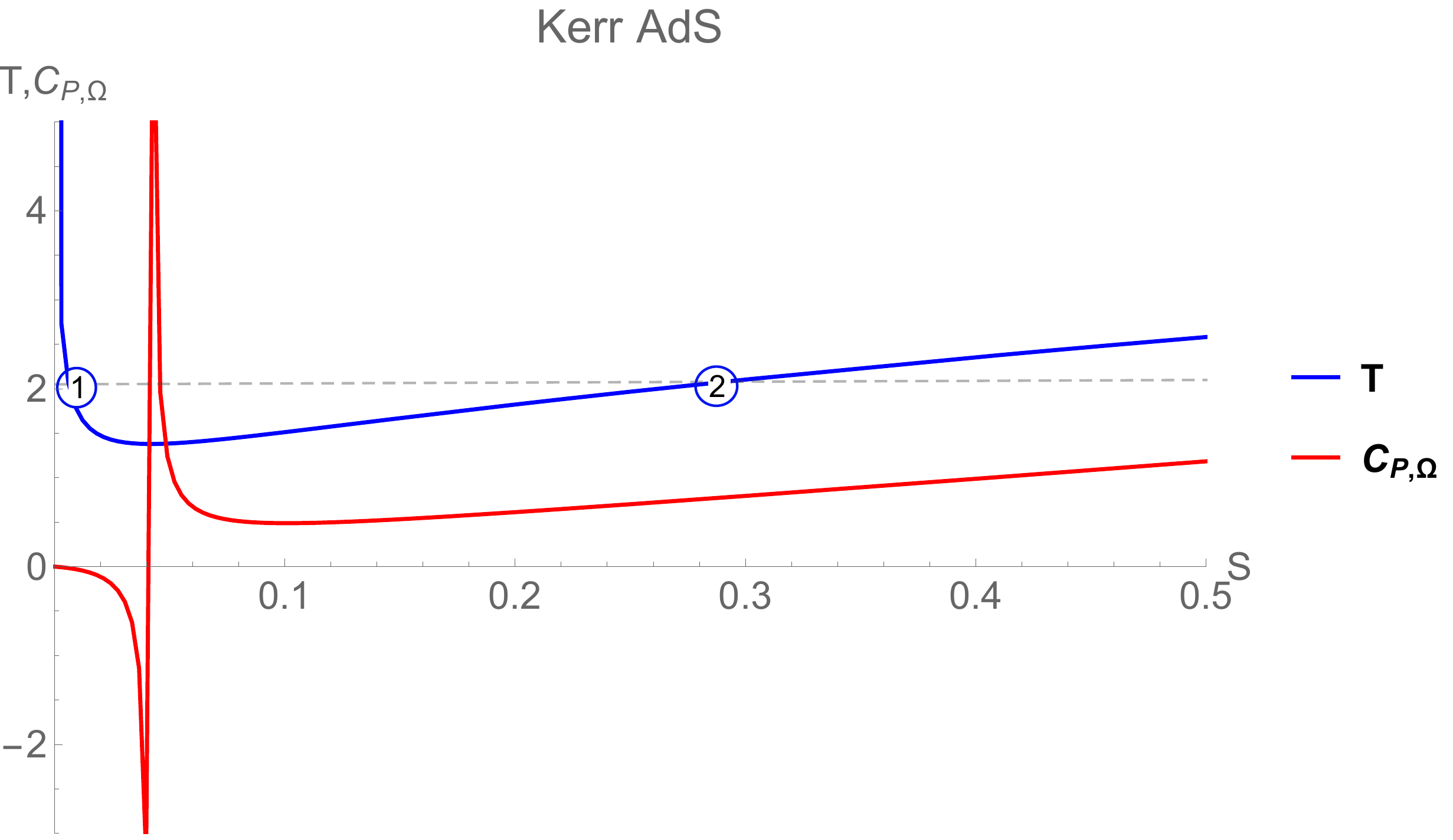}
\caption{Temperature and specific heat ($C_{P,\Omega}$) as a function of entropy for $\Omega=0.5$, showing that the small black hole is unstable whereas, the large black hole is stable.}
\label{t_and_Cp}
\end{center}
\end{figure}
which upon solving for \(S\) gives two solutions,
\begin{equation}
S=\frac{2 \pi  \left(\pm 4 \sqrt{2} \pi  \sqrt{-16 \pi  P T^2+3 T^2 \Omega ^2+8 \pi ^2 T^4}-16 \pi
   P+16 \pi ^2 T^2+3 \Omega ^2\right)}{256 \pi ^2 P^2-96 \pi  P \Omega ^2+9 \Omega ^4}.
\end{equation}
For computing \(D\), one requires to compute the specific heats and other response functions. Their expressions are found to be of the following form in \(S\), \(P\) and \(\Omega\) variables,
\begin{equation}
C_{P,\Omega}= C_{V,\Omega}+\frac{2 S \left(2 \pi  (8 P S+1)-3 S \Omega ^2\right)}{2 \pi  (8 P S-1)-3 S \Omega ^2},
\end{equation}
\begin{equation}
\alpha_V = -\frac{24 \pi ^{3/2} \sqrt{S}}{-16 \pi  P S+3 S \Omega ^2+2 \pi },
\end{equation}
\begin{equation}
\kappa_{T,\Omega} = -\frac{48 \pi  S}{-16 \pi  P S+3 S \Omega ^2+2 \pi }.
\end{equation}
The response functions \(C_{P,\Omega}\), \(\alpha_V\) and \(\kappa_{T,\Omega}\) which are specific to the \((\Omega,P,T)\)-ensemble are all found to diverge at $S=0.04$, which is the equilibrium point where new black hole phases appear as seen from figure-(\ref{t_and_Cp}). Since the thermodynamic discussion is identical to the case of charged black holes in AdS, we do not elaborate on it further here. On the other hand, the specific heat of the \((J,P,T)\)-ensemble, which reads,
\begin{equation}
C_{P,J}= C_{V,J}+\frac{2 S \left(2 \pi  (8 P S+1)-3 S \Omega ^2\right)}{2 \pi  (8 P S-1)+15 S \Omega
   ^2}
\end{equation}
diverges at the point corresponding to the critical point of the \((J,P,T)\)-ensemble which is distinct from that of the \((\Omega,P,T)\)-ensemble and can be found out from the conditions,
\begin{equation}
  \bigg(\frac{\partial T}{\partial S}\bigg)_{P,J} = 0, \hspace{5mm} \bigg(\frac{\partial^2 T}{\partial S^2}\bigg)_{P,J} = 0.
\end{equation}
Thus, we find a non-equivalence of the fixed angular momentum and fixed angular velocity ensembles. Of present interest is the \((\Omega,P,T)\)-ensemble in which we compute the logarithmic entropy corrections within the small angular momentum limit. Let us emphasize on the fact that these entropy corrections are valid only at points in the thermodynamic phase space where all the response functions are well-behaved and are positive. It should be noted that the thermodynamic volume in the small angular momentum limit is not independent of entropy and hence the specific heat \(C_{V,\Omega} = 0\). Consequently, as before, we set \(C_{V,\Omega} = \epsilon\) for an infinitesimally small positive constant \(\epsilon\). The final result for \(D\) in this small angular momentum limit is therefore,
\begin{equation}
D_{J \approx 0}=\frac{\epsilon S^2 \left(2 \pi  (8 P S+1)-3 S \Omega ^2\right)^4}{128 \pi ^7 \left(2 \pi  (8 P S-1)-3 S
   \Omega ^2\right)}.
\end{equation}
Substituting the expression for \(D_{J \approx 0}\) into eqn (\ref{entropycorrected}) one obtains the following expression (ignoring some constants),
\begin{equation}
  \mathcal{S}_{J \approx 0} = S_0 -  \ln S_0 - 2 \ln \bigg[2 \pi  (8 P S_0+1)-3 S_0 \Omega ^2 \bigg] + \frac{1}{2} \ln \bigg[2 \pi  (8 P S-1)-3 S  \Omega ^2 \bigg] + \cdots .
\end{equation} Thus, the leading order corrections include a term of the form \(-k\ln S_0\) with coefficient \(k = +1\) where \(S_0\) is given by eqn (\ref{BHS}).

\section{Conclusions}\label{conclusions}
Although the $(\mu,P,T)$-ensemble is not conventionally used to describe typical thermodynamic systems, it turns out to have applications to a wide range of systems where non-additivity plays a fundamental role~\cite{Hill1,Ruffo1}. In particular, for the case of black holes in AdS, the quasi-homogeneity of the black hole mass ensures that the system can attain equilibrium in the open ensemble and allows the discussion of thermodynamics and phase transitions in a consistent way. For a general thermodynamic system, we computed the exact analytic form of fluctuations in energy, volume and number of particles (or conserved charges for black holes) in the open or $(\mu,P,T)$-ensemble, and their expressions are summarized in table-(\ref{table1}). This complements the earlier works where fluctuations of thermodynamic quantities for general black holes were studied in~\cite{Chamblin:1999hg}, where pressure and volume terms were absent. The expressions of the elements of the covariance matrix summarizing the fluctuations were then used to obtain general logarithmic corrections to the entropy of charged as well as rotating black holes anti-de Sitter spacetimes. The current study complements our recent computations of logarithmic corrections to black hole entropy in the isothermal-isobaric ensemble \cite{Ghoshentropy}, which were in turn a follow up of earlier works where thermal fluctuations and/or fluctuations from charges were considered \cite{parthasarathi2,Mahapatra:2011si}. It is interesting to note that the leading term in the logarithmic entropy corrections is found to have the general structure \cite{parthasarathi,Ghoshentropy} of \(-k \ln S_0\), where \(S_0\) is the Bekenstein-Hawking entropy. From the inspection of various results we notice that,
\begin{equation}\label{Dinspec}
  D = \left(C_{V,\mu}T^2 \right) \left( T V \kappa_{T,\mu} \right) \left(T \chi_{T,P} \right)\, \frac{C_{P,N}}{C_{P,\mu}}  \, \, 
\end{equation}
gives the same results when plugged in eqn (\ref{entropycorrected}). In fact, (\ref{Dinspec}) vanishes identically when computed for $C_{V,\mu}=0$. In order to get a non-zero result one can normalize \(D\) by taking \(C_{V,\mu}\) to be an infinitesimal positive constant and then taking the $C_{V,\mu} \rightarrow 0$ limit of $D \rightarrow D/C_{V,\mu}$.

\section*{Acknowledgements}
A.G. gratefully acknowledges the financial support received from the Ministry of Education (MoE), Government of India in the form of a Prime Minister's Research Fellowship. S.M. thanks Goutam Tripathy for helpful discussions. C.B. thanks the Science and Engineering Research Board (DST), Government of India, through MATRICS (Mathematical Research Impact Centric Support) grant no. MTR/2020/000135. We thank the anonymous referee for helpful suggestions.

\appendix

\section{Computing the covariances}\label{appendixa}
\numberwithin{equation}{section}

The elements of the covariance matrix are defined as, 
\begin{equation}\label{Dija}
D_{ij} = \langle(\Delta y_i)(\Delta y_j)\rangle = \frac{\partial^2 S}{\partial x^i \partial x^j}\bigg|_{\rm Equilibrium}.
\end{equation} Here \(x^1 = \beta\), \(x^2 = \beta P\) and \(x^3 = \beta \mu\). It will be understood that these derivatives are computed at thermodynamic equilibrium, i.e. at \(\beta_0\), \((\beta P)_0\) and \((\beta \mu)_0\). The entropy here is defined as \(S = \ln \mathcal{Z} + \beta E + \beta PV - \beta \mu N\) with the saddle point equations,
\begin{eqnarray}
\bigg(\frac{\partial \ln \mathcal{Z}(\beta,\beta P,\beta \mu)}{\partial \beta}\bigg)_{\beta_0,(\beta P)_0,(\beta \mu)_0} &=& - E, \\
\bigg(\frac{\partial \ln \mathcal{Z}(\beta,\beta P,\beta \mu)}{\partial (\beta P)}\bigg)_{\beta_0,(\beta P)_0,(\beta \mu)_0} &=& - V, \\
\bigg(\frac{\partial \ln \mathcal{Z}(\beta,\beta P,\beta \mu)}{\partial (\beta \mu)}\bigg)_{\beta_0,(\beta P)_0,(\beta \mu)_0} &=& N.
\end{eqnarray}
Therefore, 
\begin{eqnarray}
\langle (\Delta E)^2\rangle = \bigg(\frac{\partial^2 S(\beta,\beta P,\beta \mu)}{\partial \beta^2}\bigg)_{\beta_0,(\beta P)_0,(\beta \mu)_0} &=& - \frac{\partial E}{\partial \beta}\bigg|_{\beta P, \beta \mu} , \label{EE} \\
\langle (\Delta E)(\Delta V)\rangle = \bigg(\frac{\partial^2 S(\beta,\beta P,\beta \mu)}{\partial \beta \partial (\beta P)}\bigg)_{\beta_0,(\beta P)_0,(\beta \mu)_0} &=& - \frac{\partial V}{\partial \beta}\bigg|_{\beta P, \beta \mu} , \label{EV} \\
 \langle(\Delta E)(\Delta N)\rangle = \bigg(\frac{\partial^2 S(\beta,\beta P,\beta \mu)}{\partial \beta \partial (\beta \mu)}\bigg)_{\beta_0,(\beta P)_0,(\beta \mu)_0} &=& \frac{\partial N}{\partial \beta}\bigg|_{\beta P, \beta \mu}. \label{EN}
\end{eqnarray}
In the above equations, the derivatives with respect \(\beta\) are taken at fixed \(\beta P\) and \(\beta \mu\). We can straightforwardly convert these derivatives to those at fixed \(\beta\), \(P\) or \(\mu\) giving, 
\begin{eqnarray}
\frac{\partial E}{\partial \beta}\bigg|_{\beta P, \beta \mu} &=&  \frac{\partial E}{\partial \beta}\bigg|_{P,\mu} - \frac{P}{\beta} \frac{\partial E}{\partial P}\bigg|_{\beta, \mu} - \frac{\mu}{\beta} \frac{\partial E}{\partial \mu}\bigg|_{\beta, P}
\end{eqnarray} and similarly for \(V\) and \(N\). Thus, eqns (\ref{EE})-(\ref{EN}) can be re-written as, 
\begin{eqnarray}
\langle (\Delta E)^2\rangle &=& -  \frac{\partial E}{\partial \beta}\bigg|_{P,\mu} + \frac{P}{\beta} \frac{\partial E}{\partial P}\bigg|_{\beta, \mu} + \frac{\mu}{\beta} \frac{\partial E}{\partial \mu}\bigg|_{\beta, P}  ,\\
\langle (\Delta E)(\Delta V)\rangle &=&  -\frac{\partial V}{\partial \beta}\bigg|_{P,\mu} + \frac{P}{\beta} \frac{\partial V}{\partial P}\bigg|_{\beta, \mu} - \frac{\mu}{\beta} \frac{\partial V}{\partial \mu}\bigg|_{\beta, P} , \\
\langle (\Delta E)(\Delta N)\rangle &=&  \frac{\partial N}{\partial \beta}\bigg|_{P,\mu} - \frac{P}{\beta} \frac{\partial N}{\partial P}\bigg|_{\beta, \mu} - \frac{\mu}{\beta} \frac{\partial N}{\partial \mu}\bigg|_{\beta, P}
\end{eqnarray} which can be re-arranged to give the final expressions for the covariances \(\langle(\Delta E)^2\rangle\), \(\langle(\Delta E)(\Delta V)\rangle\) and \(\langle(\Delta E)(\Delta N)\rangle\) listed in table-(1). In an exactly similar manner, one can obtain the remaining three elements of the covariance matrix, i.e. \(\langle (\Delta V)^2\rangle\), \( \langle (\Delta N)^2\rangle\) and \(\langle (\Delta V)(\Delta N)\rangle\). From eqn (\ref{Dija}), they are defined as, 
\begin{eqnarray}
\langle (\Delta V)^2\rangle = \bigg(\frac{\partial^2 S(\beta,\beta P,\beta \mu)}{\partial (\beta P)^2}\bigg)_{\beta_0,(\beta P)_0,(\beta \mu)_0} &=& - \frac{\partial V}{\partial (\beta P)}\bigg|_{\beta, \beta \mu} , \label{VV}   \\
\langle (\Delta N)^2\rangle  = \bigg(\frac{\partial^2 S(\beta,\beta P,\beta \mu)}{\partial (\beta \mu)^2}\bigg)_{\beta_0,(\beta P)_0,(\beta \mu)_0} &=&  \frac{\partial N}{\partial (\beta \mu)}\bigg|_{\beta, \beta P} , \label{NN} \\
\langle (\Delta V)(\Delta N)\rangle  = \bigg(\frac{\partial^2 S(\beta,\beta P,\beta \mu)}{\partial (\beta P) \partial (\beta \mu)}\bigg)_{\beta_0,(\beta P)_0,(\beta \mu)_0} &=& - \frac{\partial V}{\partial (\beta \mu)}\bigg|_{\beta, \beta P} . \label{NV}
\end{eqnarray}
Consider eqn (\ref{VV}). Since, the derivative on the left hand side involves a fixed temperature, we may write, 
\begin{equation}
\langle (\Delta V)^2\rangle = - \frac{1}{\beta} \frac{\partial V}{\partial P}\bigg|_{\beta, \beta \mu} = - \frac{1}{\beta} \frac{\partial V}{\partial P}\bigg|_{\beta, \mu}
\end{equation} where, in the last equality we have used the fact that if \(\beta\) and \(\beta \mu\) both are fixed parameters, then \(\mu\) individually must be fixed. The above expression straightforwardly gives the expression for \(\langle (\Delta V)^2\rangle\) listed in table-(1). Similarly, using the same arguments, it is easy to check that eqn (\ref{NV}) reproduces the correct expression for \(\langle (\Delta V)(\Delta N)\rangle\) stated earlier. Finally, consider eqn (\ref{NN}) which can be re-written invoking similar arguments (due to constancy of \(\beta\)) as above in the following form, 
\begin{equation}
\langle (\Delta N)^2\rangle  = \frac{1}{\beta} \frac{\partial N}{\partial \mu}\bigg|_{\beta, P}
\end{equation} which straightforwardly gives the expression for \(\langle (\Delta N)^2\rangle\) listed in table-(1).

\section{Maxwell's relations}\label{appendixb}
\numberwithin{equation}{section}
We have six thermodynamic variables \(\{T,S,P,V,\mu,N\}\) and four thermodynamic energy functions, namely \(M\) (which is the enthalpy), the internal energy \(E = M - PV\), the energy function of the grand canonical ensemble, i.e. \(U = M - PV - \mu N\) and energy function of the \((\mu,P,T)\)-ensemble, i.e. \(\mathcal{E} = M - \mu N\). They satisfy the following first laws,
\begin{equation}
  dM = TdS + VdP + \mu dN, \hspace{5mm} dE = TdS - PdV + \mu dN,
\end{equation}
\begin{equation}
  dU = TdS - PdV - N d\mu, \hspace{5mm} d\mathcal{E} = TdS + VdP - N d\mu.
\end{equation}
Furthermore, one can define the free energy functions \(G = M - TS\), \(F = E - TS\), \(\Psi = U - TS\) and \(\mathcal{F} = \mathcal{E} - TS\) which respectively satisfy,
\begin{equation}
  dG = -SdT + VdP + \mu dN, \hspace{5mm} dF = -SdT - PdV + \mu dN,
\end{equation}
\begin{equation}
  d\Psi = -S dT - PdV - N d\mu, \hspace{5mm} d\mathcal{F} = -S dT + VdP - N d\mu.
\end{equation}
Using these relations, one can straightforwardly derive Maxwell-like relations for the present case. Let us first note that upon fixing each of the six thermodynamic variables, it is possible to obtain four Maxwell-like relations. Therefore, here we have \(4 \times 6 = 24\) distinct Maxwell's relations which are listed below.
\begin{itemize}
   \item \textbf{At fixed potential:} \begin{equation}
                                                     \bigg(\frac{\partial T}{\partial P}\bigg)_{S,\mu} = \bigg(\frac{\partial V}{\partial S}\bigg)_{P,\mu}, \qquad  \bigg(\frac{\partial S}{\partial P}\bigg)_{T,\mu} = -\bigg(\frac{\partial V}{\partial T}\bigg)_{P,\mu},
                                                \end{equation}
                                                \begin{equation}
                                                  \bigg(\frac{\partial P}{\partial T}\bigg)_{V,\mu} = \bigg(\frac{\partial S}{\partial V}\bigg)_{T,\mu}, \qquad  \bigg(\frac{\partial P}{\partial S}\bigg)_{V,\mu} = -\bigg(\frac{\partial T}{\partial V}\bigg)_{S,\mu}.
                                                \end{equation}
                                                \item \textbf{At fixed charge:} \begin{equation}
                                                     \bigg(\frac{\partial T}{\partial P}\bigg)_{S,N} = \bigg(\frac{\partial V}{\partial S}\bigg)_{P,N}, \qquad  \bigg(\frac{\partial S}{\partial P}\bigg)_{T,N} = -\bigg(\frac{\partial V}{\partial T}\bigg)_{P,N},
                                                \end{equation}
                                                \begin{equation}
                                                  \bigg(\frac{\partial P}{\partial T}\bigg)_{V,N} = \bigg(\frac{\partial S}{\partial V}\bigg)_{T,N}, \qquad  \bigg(\frac{\partial P}{\partial S}\bigg)_{V,N} = -\bigg(\frac{\partial T}{\partial V}\bigg)_{S,N}.
                                                \end{equation}
   \item   \textbf{At fixed pressure:}
                                                \begin{equation}
                                                  \bigg(\frac{\partial T}{\partial \mu}\bigg)_{S,P} = -\bigg(\frac{\partial N}{\partial S}\bigg)_{\mu,P}, \qquad  \bigg(\frac{\partial N}{\partial T}\bigg)_{\mu,P} = \bigg(\frac{\partial S}{\partial \mu}\bigg)_{T,P},
                                                \end{equation}
                                                \begin{equation}
                                                  \bigg(\frac{\partial S}{\partial N}\bigg)_{T,P} = -\bigg(\frac{\partial \mu}{\partial T}\bigg)_{N,P}, \qquad  \bigg(\frac{\partial T}{\partial N}\bigg)_{S,P} = \bigg(\frac{\partial \mu}{\partial S}\bigg)_{N,P}.
                                                \end{equation}
    \item   \textbf{At fixed volume:}
                                                \begin{equation}
                                                  \bigg(\frac{\partial T}{\partial \mu}\bigg)_{S,V} = -\bigg(\frac{\partial N}{\partial S}\bigg)_{\mu,V}, \qquad  \bigg(\frac{\partial N}{\partial T}\bigg)_{\mu,V} = \bigg(\frac{\partial S}{\partial \mu}\bigg)_{T,V},
                                                \end{equation}
                                                \begin{equation}
                                                  \bigg(\frac{\partial S}{\partial N}\bigg)_{T,V} = -\bigg(\frac{\partial \mu}{\partial T}\bigg)_{N,V}, \qquad  \bigg(\frac{\partial T}{\partial N}\bigg)_{S,V} = \bigg(\frac{\partial \mu}{\partial S}\bigg)_{N,V}.
                                                \end{equation}
    \item   \textbf{At fixed entropy:}
                                                \begin{equation}
                                                  \bigg(\frac{\partial V}{\partial \mu}\bigg)_{P,S} = -\bigg(\frac{\partial N}{\partial P}\bigg)_{\mu,S}, \qquad \bigg(\frac{\partial N}{\partial V}\bigg)_{\mu,S} = \bigg(\frac{\partial P}{\partial \mu}\bigg)_{V,S},
                                                \end{equation}
                                                \begin{equation}
                                                  \bigg(\frac{\partial P}{\partial N}\bigg)_{V,S} = -\bigg(\frac{\partial \mu}{\partial V}\bigg)_{N,S}, \qquad    \bigg(\frac{\partial \mu}{\partial P}\bigg)_{N,S} = \bigg(\frac{\partial V}{\partial N}\bigg)_{P,S}.
                                                \end{equation}
    \item   \textbf{At fixed temperature:}
                                                \begin{equation}
                                                  \bigg(\frac{\partial V}{\partial \mu}\bigg)_{P,T} = -\bigg(\frac{\partial N}{\partial P}\bigg)_{\mu,T}, \qquad  \bigg(\frac{\partial N}{\partial V}\bigg)_{\mu,T} = \bigg(\frac{\partial P}{\partial \mu}\bigg)_{V,T},
                                                \end{equation}
                                                \begin{equation}
                                                  \bigg(\frac{\partial P}{\partial N}\bigg)_{V,T} = -\bigg(\frac{\partial \mu}{\partial V}\bigg)_{N,T}, \qquad \bigg(\frac{\partial \mu}{\partial P}\bigg)_{N,T} = \bigg(\frac{\partial V}{\partial N}\bigg)_{P,T}.
                                                \end{equation}

 \end{itemize}
Such relations for charged or rotating black holes can be obtained by making appropriate identifications.

\end{document}